\documentclass[preprint,authoryear,12pt]{elsarticle}

\newcommand{\aap}{    {\it A \& A }}
\newcommand{\solphys}{    {\it Sol. Phys. }}
\newcommand{\jgr}{    {\it JGR }}
\newcommand{\ssr}{    {\it SSR  }}
\newcommand{\apjl}{    {\it ApJL  }}
\newcommand{\apj}{    {\it ApJ  }}

\newcommand{\mnras}{    {\it MNRAS  }}

\usepackage{graphicx}
\usepackage{epsfig}
\usepackage{times}
\usepackage{natbib}
\usepackage{url}
\usepackage{color}
\usepackage{amssymb,amsmath}
\usepackage{amssymb}

\begin{document}
\begin{frontmatter}

\title{A Study of a long duration B9 flare-CME event and associated piston-driven shock}

\author{R. Chandra\corref{cor}}
\address{Department of Physics, DSB Campus, Kumaun University, Nainital -- 263 001, India}
\cortext[cor]{Corresponding author}
\ead{rchandra.ntl@gmail.com}

\author{P. F. Chen}
\address{School of Astronomy \& Space Science, Nanjing University, Nanjing 210023, China}

\author{A. Fulara}
\address{Department of Physics, DSB Campus, Kumaun University, Nainital -- 263 001, India}

\author{A. K. Srivastava}
\address{Department of Physics, Indian Institute of Technology (BHU), Varanasi -- 221 005, India}

\author{W. Uddin}
\address{Aryabhatta Research Institute of Observational Sciences, Manora Peak, Nainital -- 263 001, India}

\begin{abstract} We present and discuss here the observations of a small long duration GOES B-class flare 
        associated with a quiescent 
	filament eruption, a global EUV wave and a CME on 2011 May 11. The event was well observed by the Solar 
	Dynamics Observatory (SDO), GONG H$\alpha$, STEREO and HiRAS spectrograph. 
        As the filament erupted, ahead of 
	the filament we observed the propagation of EIT wave fronts, as well as two flare ribbons on both sides of 
	the polarity inversion line (PIL) on the solar surface. The observations show the co-existence of two types 
	of EUV waves, i.e., a fast and a slow one. A type II radio burst with {\bf up to} the 
        third harmonic component was 
	also associated with this event. The evolution of photospheric magnetic field showed flux emergence and 
	cancellation at the filament site before its eruption.
\end{abstract}


\end{frontmatter}

\section{Introduction}

Solar flares are the most energetic phenomenon near the solar surface and occasionally they are accompanied by 
filament (or prominence) eruptions and coronal mass ejections (CMEs) \citep{Chen11,Fletcher11,Joshi12,Benz17}. 
The association of filament eruptions with solar flares varies from small GOES B-class flares to very 
large GOES X-class flares.  
Filaments are dense cool plasma materials suspended in the hot corona \citep{Parenti14}. They are often  
visible {\bf at} chromospheric and coronal heights. Observations indicate that they are along the polarity 
inversion line (PIL). The processes related to the flare occurrence, the formation of flare ribbons, and the eruption of 
filament are well explained by standard CSHKP model \citep{Carmichael64,Sturrock66,Hirayama74,Kopp76}. 

Sometimes solar flares along with erupting filaments are accompanied by 
globally propagating waves, known as extreme ultraviolet (EUV) waves. 
EUV waves were first observed by the Extreme Ultraviolet Imaging Telescope(EIT, \cite{Del95}) on-board the SOHO spacecraft, and were thus 
historically named as EIT waves \citep{Thompson99,Thompson00}.
These EUV waves can propagate to long distances on the solar disk with a speed of about 
170--350 $km s^{-1}$ \citep{Thompson00}. Now, with the high cadence SDO data, our knowledge on EUV waves is 
enhanced considerably. EIT waves were initially considered as the coronal counterpart of high 
speed Chromospheric Moreton waves \citep{Moreton1960}, although it was also noticed that the EIT wave speeds are several times smaller than those of Moreton waves.. Later on, \citet{Warmuth04} and \citet{Vrsnak16}
explained that the difference in the velocities can be attributed to the deceleration
 of coronal waves. 

The interpretations for the EUV waves include wave and non-wave models. The wave models interpret EUV waves as fast-mode MHD waves \citep{Thompson99, Wang2000}. The observations like reflection, refraction, 
transmission seem to support the wave nature of EUV waves \citep{Olmedo12}.
The discovery of stationary fronts \citep{Delannee99} of EUV waves challenged the wave nature of EUV waves. 
Later on several other models have been proposed which include the magnetic field-line stretching model \citep{Chen02,Chen05},  
the successive reconnection model \citep{Attrill07}, the slow-mode wave model 
\citep{Wills07,Wang09}, and the current shell model \citep{Dela08}. The magnetic field line stretching model further  
proposed that there should be a fast-mode wave ahead of the slow EUV wave. It is believed in the above discussed observations 
and models that the slow wave stops at the magnetic quasi-separatrix layers (QSLs) and forms stationary fronts. 
However, very recently \cite{Chandra16}
found the observation of stationary fronts associated with the fast component of EUV waves and their location also lies at the QSLs.
Encouraged by this observational finding, \cite{Chen16} did a numerical simulation of the interaction of a fast mode MHD wave and a magnetic QSL. Their numerical results showed that the fast-mode MHD wave does generate a stationary front once passing through a 
magnetic QSL. Their study suggested that some part of this fast-mode wave is converted to a slow mode wave which 
gets trapped and forms a stationary front. Recently, this type of stationary waves is confirmed by the studies of \cite{Yuan16,Sri16}, and 
\cite{Zong17}.

Solar flares, filament eruptions, and EUV waves are sometimes associated with type II radio bursts. Type II radio bursts
 are the signature of shock waves propagating in the corona. Type II radio bursts are slowly drifting radio emission from high to 
low frequencies \citep{Wild50}. Both the EUV waves and type II radio bursts are often
associated with CMEs. These type II radio bursts are believed to be triggered by either a CME
\citep{Cliver99, Gopal2001} or by a blast wave which gets created by a flare \citep{Uchida1974, Hudson04}.
It is difficult to determine whether a shock is ignited by a CME or a flare. \cite{Vasanth11} studied the characteristics
 of type II radio bursts associated with flares and CMEs and concluded that some parts of the high frequency shocks are initiated by flares,
 whereas low frequency type II bursts are related to the shocks driven by CMEs. 
Also, \citet{Gopal99} did a study of type II radio bursts and CMEs and suggested that at the height of 
minimum Alfv\'en speed a type II radio burst starts and the end time will depend on the relative 
variation of the CME speed and the Alfv\'en speed of the background corona.

The aim of this paper is to study the flare, filament eruption and their association with EUV waves, CME and type II radio bursts on 2011 May 11. The paper is organized as follows: Section 2 describes the observations and in Section 3, we present the analysis and results of the study. Finally in Section 4, we summarize our study. 

\section{Observations}
\label{obs}

For the current study, we used the data from the following sources:

\begin{itemize}
\item{\bf SDO/AIA and HMI data:}  
The Atmospheric Imaging Assembly (AIA, \citep{Lemen12}) on board Solar Dynamic Observatory (SDO, \cite{Pesnell12}) observes the full Sun with different filters in EUV and UV spectral lines. The cadence is 12 sec and the pixel size is  
0.6 arcsec. For this study, we used the AIA 171 \AA, 193 \AA, 211 \AA, and 335 \AA\ data. In order investigate the 
magnetic causes of the filament eruption and the associated flare, we used the data from the Heliospheric Magnetic 
Images (HMI,\citep{Scherrer12}) observations. HMI observes the photospheric magnetic field of the  Sun with a cadence of 45 sec and spatial resolution 1$''$ (i.e., the pixel size is 0.5$''$). 

\item{\bf NSO/GONG data:} 
For the chromospheric observations of the filament eruption and flare, we used the H$\alpha$ data from the National 
Solar Observatory (NSO)/ Global Oscillation Network Group (GONG,\citep{Harvey11}) instrument. GONG observes the full Sun in H$\alpha$ with a cadence of 1 min. The spatial resolution of the GONG data is 2$''$ (i.e., the pixel size is 1$''$). 

\item{\bf STEREO, LASCO, and Radio spectrograph data:} 
To look into the associated CME with the filament eruption, we used the COR1, COR2 \citep{Kaiser08} and 
LASCO \citep{Brueckner95} CME data. For the radio analysis, we used the HiRAS spectrograph data.
\end{itemize}

\section{Analysis and Results}
\label{analysis}

\subsection{Overview of the event and EUV wave}
\label{overview}

On 2011 May 11 the filament under study was located between the active regions \textit{NOAA 11207} and \textit{NOAA 11205} at \textit{N20W60} on the solar disk. Earlier, this filament eruption was studied by \citet{Chandra16} and \citet{Grechnev15}. \citet{Chandra16} reported that the eruption was associated with two global EUV waves, which were predicted by \citet{Chen02}. In their study, apart from the traditional stationary EUV front which propagates slowly in the early phase and then stops a magnetic separatrix, they reported for the first time that the coronal fast-mode EUV wave also generates a stationary front close 
to another QSL. They interpretated that the magnetic structure near the stationary front could be a magnetic valley, where 
a part of the fast-mode wave is trapped. Later on, \citet{Chen16} did a 2D numerical simulation and found that the 
stationary front results from the fast coronal wave and is located close to a magnetic QSL, which are consistent 
with the observations of \citet{Chandra16}. Their analysis indicates that at the place where plasma beta is equal to unity, a part of the fast-mode wave is converted into a slow-mode MHD wave. The converted slow MHD wave cannot cross the magnetic loops and forms a stationary wave front. It has been known that when a fast-mode wave enters the site with weak magnetic field where the Alfv\'en speed is similar to the sound speed, a part of the fast-mode wave can be converted to a slow-mode wave \citep{Cally05, mcla06}. 
 
\begin{figure*}
\centering
\includegraphics[width=0.8\textwidth,clip=]{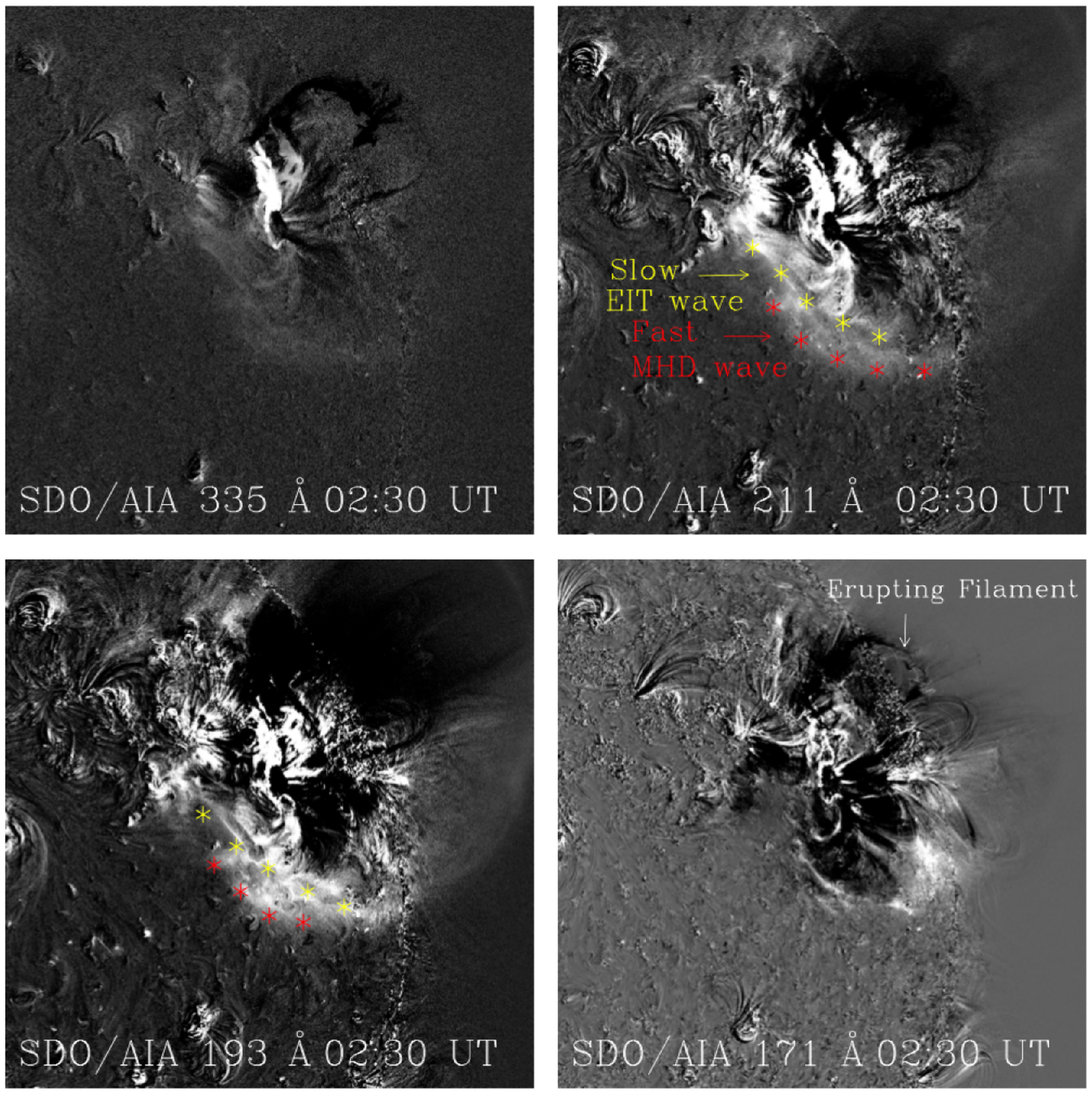}
\caption{A snapshot of the slow and fast EUV waves visible in different AIA wavelengths. The fast and slow waves are indicated by red and yellow asterisks, respectively.}
\label{wave}
\end{figure*}

The evolution of the fast and slow EUV waves shows multi-wavelength characteristics. Figure \ref{wave} shows both the 
waves appearing in the AIA 335 \AA, 211 \AA, 193 \AA, and 171 \AA\  channels. As we can see, the waves are promptly visible in AIA 193 \AA, which is a historical wavelength where EIT waves were discovered. 
In Table 1, we present the chronology of the event.

\begin{table}[t]
\caption{Chronology of event}
\medskip
\centering
\begin{tabular}{c c}
\hline
Time in UT & Event \\
\hline 
02:10  & Onset of the filament eruption \\
02:20  & Onset of the GOES B-Class flare \\
02:23  & Onset of the EIT wave \\
02:25  & Onset of the CME  \\
02:28  & HiRAS type II radio burst appearance\\
02:28  & Shock onset time \\
02:43  & HiRAS type II third harmonic appearance\\
\hline
\end{tabular}
\end{table}

\subsection{Filament eruption and flare dynamics}
\label{filamnet}

\begin{figure*}
\centering
\includegraphics[width=1.0\textwidth,clip=]{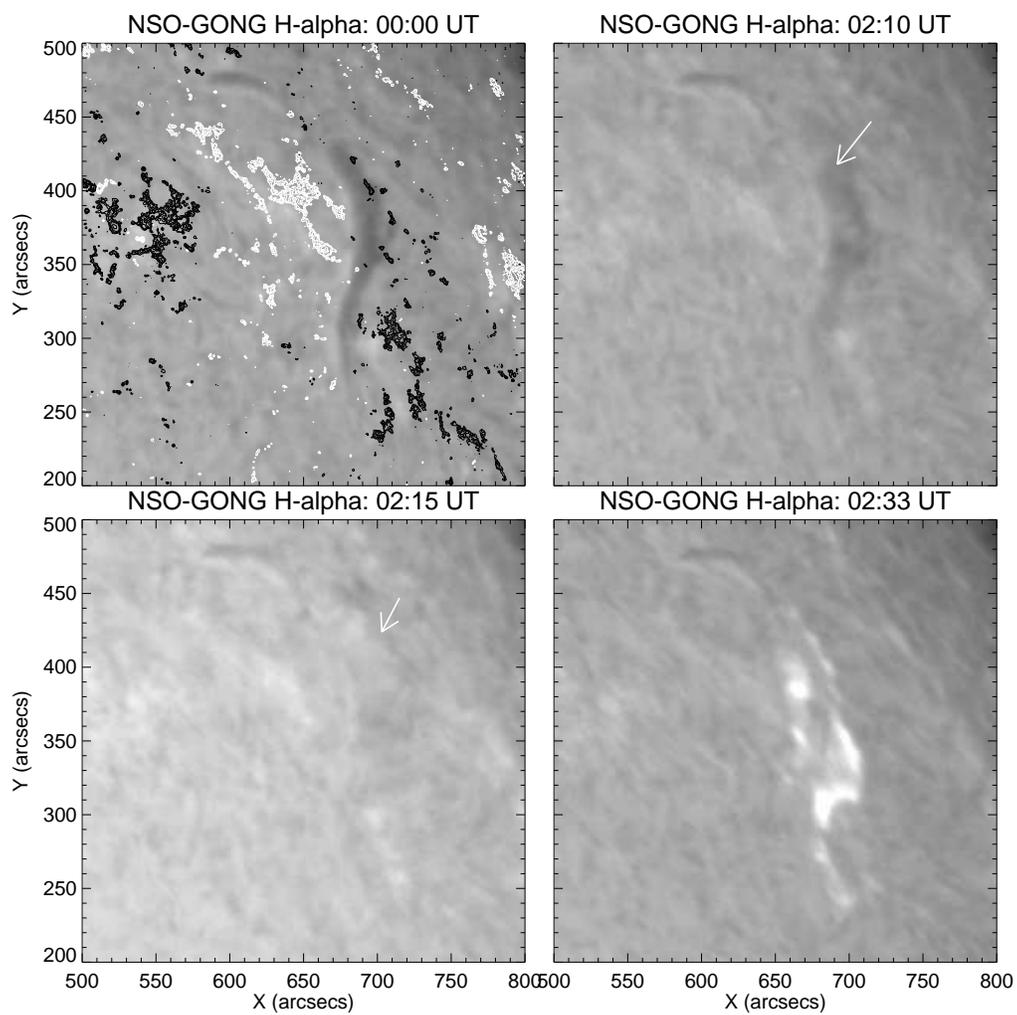}
\caption{Evolution of the filament eruption followed by a flare observed by NSO/GONG in H$\alpha$. The first image is overlaid by HMI magnetic field contours, where white/black contours correspond to positive/negative polarities, respectively.}
\label{hal}
\end{figure*}

On 2011 May 11 there were two end-to-end filaments, viz., northern and southern. The southern filament was very big,
 its length was around 150 Mm, while the northern filament was smaller (with a length of $\approx$60 Mm). In this study, we concentrate on the big southern filament, which erupted later on. First, we determine the handedness of this filament based on the definition of \citet{Mackay10}. We  identify the end points of the filament and the associated magnetic polarity. The northern endpoint is found to be rooted at the positive polarity, while the south end is rooted at the negative polarity. The corresponding H$\alpha$ filtergram overlaid by HMI magnetic field contours is presented in the first image of Figure \ref{hal}. Based on the the magnetic polarities of the two end-points of the filament, we found the filament has dextral chirality, i.e. left-handedness. According to \cite{Hao16}, the above-mentioned method sometimes does not work well. In order to confirm the result, we also determine the chirality of the filament by two other independent methods, i.e., filament barbs \citep{Mart08} and the drainage sites of the erupting filament \citep{Chen14}. Looking at the barbs of the filaments, we can also see the dextral barbs, which indicates that the filament has left handedness. After checking the skew of the conjugate drainage sites in EUV wavelength, we also come to the result that the filament has left handedness. The left hand handedness confirms the hemispheric rule \citet{Pevtsov95, Ouya17}.       

The filament started to rise at $\approx$02:10 UT on 2011 May 11. The eruption of the filament was recorded 
by \textit{AIA} 
in all its channels and by \textit{GONG} in H$\alpha$. As the filament was erupting, in \textit{AIA} 171 \AA\ at 02:25:13 UT, we observed the flux rope. The twist of this formed flux rope is consistent with the handedness of the filament 
visible in H$\alpha$ wavelength.

\begin{figure*}
\centering
\includegraphics[width=1.0\textwidth,clip=]{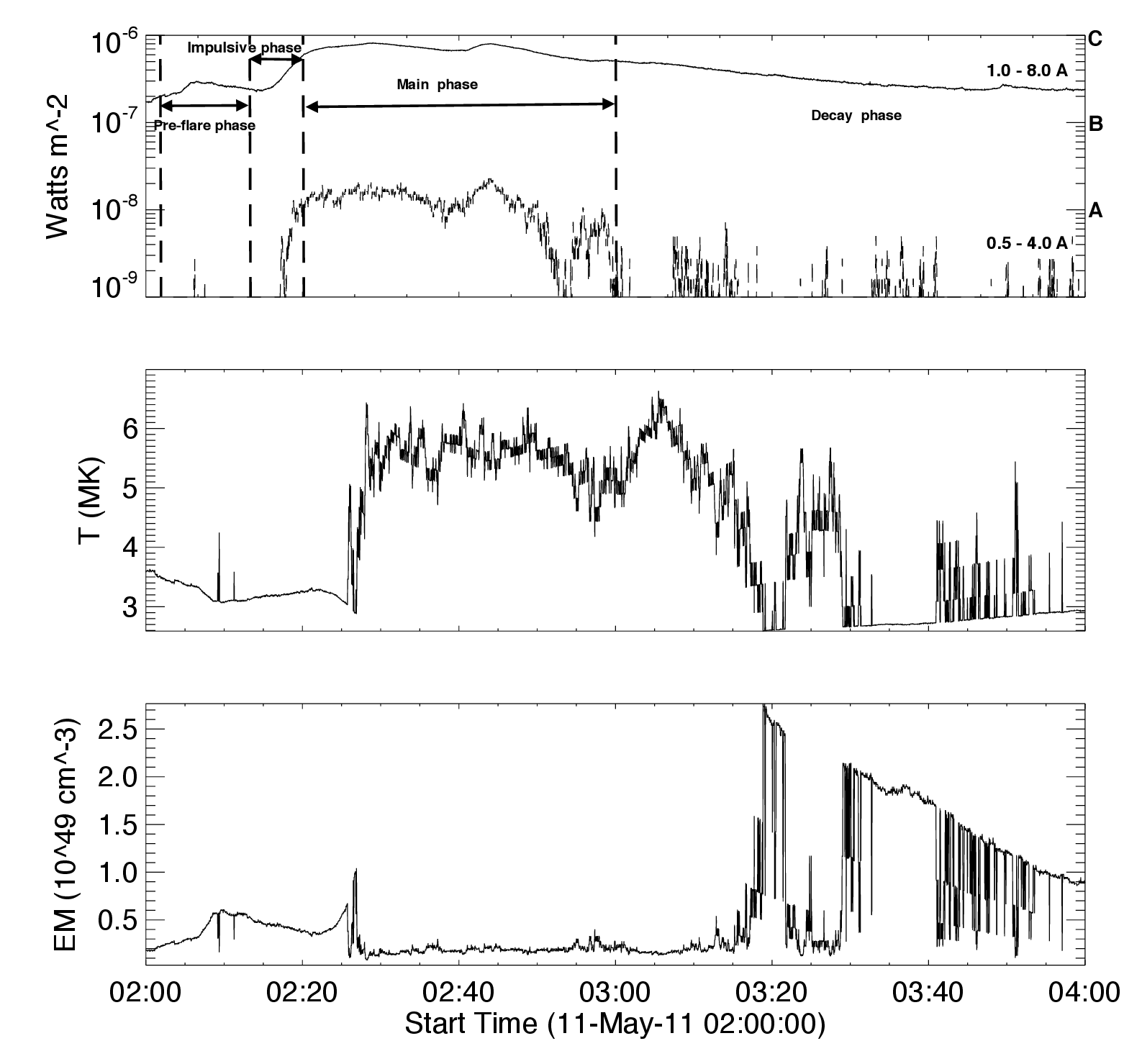}
\caption{Evolution of the soft X-ray flux of the flare observed by GOES (top) and the evolution of temperature (middle) and emission measure (bottom) .}
\label{goes}
\end{figure*}

The eruption of the filament was followed by a small \textit{GOES} B-class flare. According to \textit{GOES} 
observations, the flare was classified as B9.0-class. It was initiated around 02:20 UT, peaked 02:40 UT and decayed after 03:20 UT. 
The \textit{GOES} temporal evolution of the flare is shown in Figure \ref{goes} (top panel). In the GOES profile, we notice three peaks.
The first peak was $\approx$02:10 UT and the second and third peak were around 02:45 and 03:05 UT, respectively. The
first peak could be the pre-flare brightening as observed in the literature \citep{Chifor07,Joshi11,Arun14,Joshi16}. 
Interestingly, the time of the first peak is consistent with the start time of filament eruption.
We have also computed the temperature and emission measure during the flare using GOES data. The results of this is shown in Figure \ref{goes} 
(middle and bottom panel). The value of the temperature during the peak flare phase is around 6 MK. 
During the pre-flare phase, a small rise in the emission measure is also noticed. {\bf
We noticed the anomaly in the estimation of emission measure during the period $\sim$ 02:27 - 03:10 UT.  Since this calculation
is based on the filter-ratio method. As we discussed the flares is very small i.e. GOES B-class and the intensity is very low in GOES
channels. Therefore the low GOES intensity could be responsible for the anomaly in emission measure estimation.} 

Looking at the spatial evolution of the flare in \textit{AIA} 171, 304 \AA\  as well as in GONG/H$\alpha$, the 
flare presented two parallel ribbons. As the filament was moving up, the two ribbons started to 
separate from each other, as expected by the standard CHSKP model. The ribbons are located on the 
opposite sides of the PIL. The ribbon structures have a reverse `J' shape and 
they were shifted along the PIL. The `J'-shaped flare ribbons were first modelled by \citet{Demoulin96} and {\bf were} 
later on were confirmed in several observations \citep{Chandra09,Schrijver11,Aulanier12,Zhao16}. 
Very recently,\cite{Joshi17} reported the observations of  `J'-shaped flare ribbons in the impulsive phase of the flare.
The `J'-shaped ribbons are indicative of magnetic shear. The sign of shear depends on the shape of the ribbons: If the 
shape looks like a forward/reverse `J', this represents positive/negative twist of the erupting flux rope, 
respectively.
Therefore, in our case the {\bf twist of the erupting flux rope} is negative, which is consistent with the 
chirality of the erupting filament. The evolution of the filament eruption and the flare 
observed by \textit{AIA} and \textit{GONG} is shown in Figures \ref{wave} and \ref{hal}, respectively.

\subsection{CME and driven shock wave}
\label{dynamic}
\begin{figure*}
\centering
\includegraphics[width=1.0\textwidth,clip=]{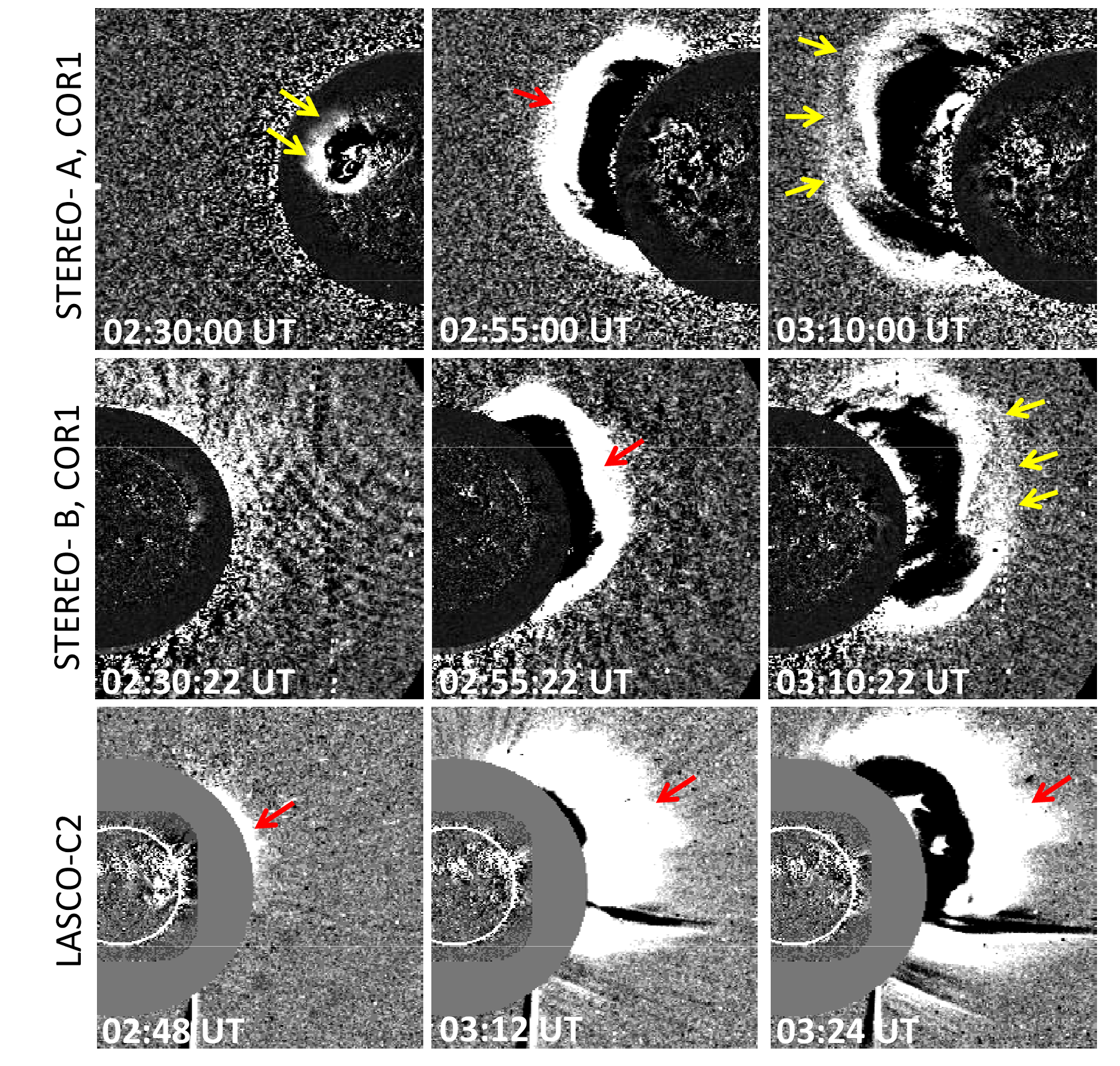}
\caption{Evolution of the associated CME observed by STEREO/COR1 (top and middle panels) and SOHO/LASCO C2 coronagraphs (bottom panel), respectively.}
\label{CME_fig}
\end{figure*}

The filament eruption was associated with a CME. The CME appeared in the LASCO
field-of-view around 02:48 UT (Figure \ref{CME_fig}, bottom panel). The CME was
a partial halo event with an angular width of 225$^{\circ}$. The speed and the
acceleration of the CME are 740 km s$^{-1}$ and 3.3 m s$^{-2}$, respectively.
The CME is observed by the COR1 coronagraph aboard the {\em STEREO} twin
spacecraft (A and B). {\em STEREO}/COR1 observes 
the inner corona with a field-of-view of 1.5--4$R_\odot$. The evolution of the
CME observed by the {\em STEREO}/COR1 coronagraph is displayed in the top and
middle panels of Figure \ref{CME_fig}. In the figure, we can see a shock wave
ahead of the CME, which is pointed out by yellow arrows, whereas the CME
leading edge is marked by red arrows. The bright core embedded in the cavity
is generally believed to be the erupting filament.

\begin{figure*}
\centering
\hspace*{-2cm}
\includegraphics[width=0.6\textwidth,angle=-90,clip=]{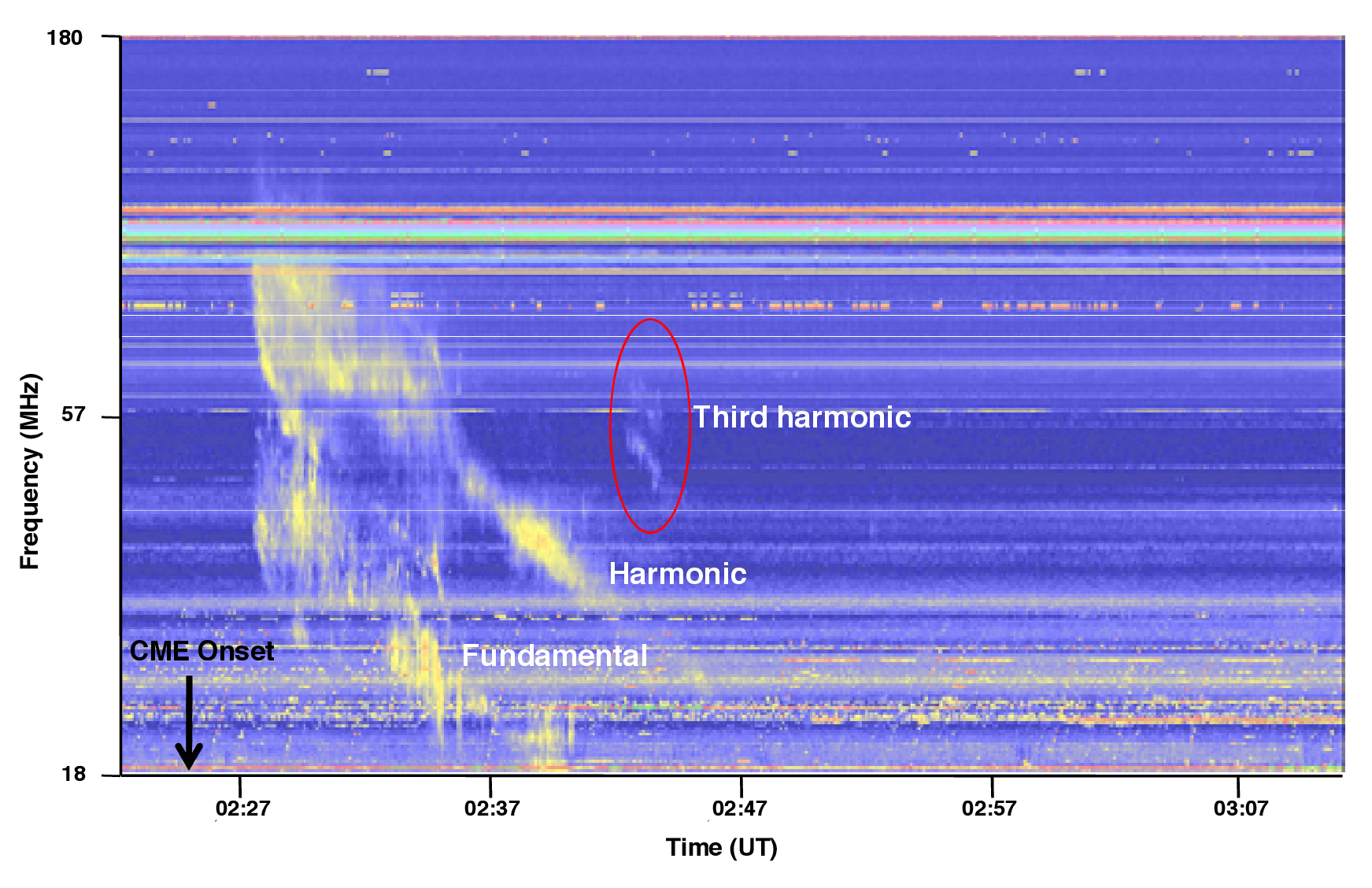}
\caption{The dynamic radio spectrum observed by HiRAS on 2011 May 11. The
fundamental and the second harmonic components of bursts are clearly visible
in the figure. The weak third harmonic is encircled.}
\label{radio}
\end{figure*}

The filament eruption is also associated with type II radio bursts. The
existence of the type II radio bursts confirms the presence of a shock.
Figure \ref{radio} displays the radio dynamic spectrum observed by the HiRAS
spectrograph. It is seen that radio bursts started from 02:28 UT, when type III
bursts and type II bursts were intermingled. Only after 02:36 UT, the type II
radio bursts stand out clearly, which clearly show the fundamental (F) and the
harmonic (H) bands. We also observed the third harmonic component, as marked by
the red ellipse in Figure \ref{radio}. The observation of the third harmonic is
unusual and has been reported in only a few observations in the past
\citep{Kliem92, Zlotnik98}. The third harmonic component is  split into two parallel
sub-bands (see Figure \ref{radio}). Such band-splitting can be seen in the F
and H bands in many events, but not in this event. The two parallel sub-bands
are the consequence of the plasma emission from the upstream and the
downstream sides of the shock wave \citep{Smerd74}. These upstream and downstream
sub-bands are called f$_U$ (upper frequency branch, UFB) and f$_L$ (lower
frequency branch, LUF), respectively \citep{Vrsnak02}.  Recently,
\citet{Zucca14} studied the eruption of the 2013 November 06 CME event, and
they also observed two parallel lanes in the harmonic component of the type II
radio bursts. They also measured the position of the type II radio source using
the Nancay Radioheliograph (NRH, \citet{Ker97}) data at different times and frequencies. They found that the type II burst
source was located above the leading edge of the CME, supporting the idea that
the shock is driven by the CME.

Following \citet{Vrsnak01,Vrsnak02}, in the case of low plasma $\beta$, we can
calculate the Alfv\'en Mach number, $M_A$, as follows:

\begin{equation}
M_A = \sqrt{\frac{X(X+5)}{2(4-X)}},
\end{equation}

\noindent
where $X = (f_U/f_L)^2$. From Figure \ref{radio}, the values of $f_U$ and
$f_L$ are 57 MHz and 45.3 MHz, respectively. The resulting Mach number is
derived to be 1.46, which is in accordance with previously reported values
\citep{Zucca14, Vrsnak02, Nindos11, Su15}.

\subsection{Evolution of photospheric magnetic field}
\label{magnetic}

\begin{figure*}
\centering
\includegraphics[width=1.0\textwidth,clip=]{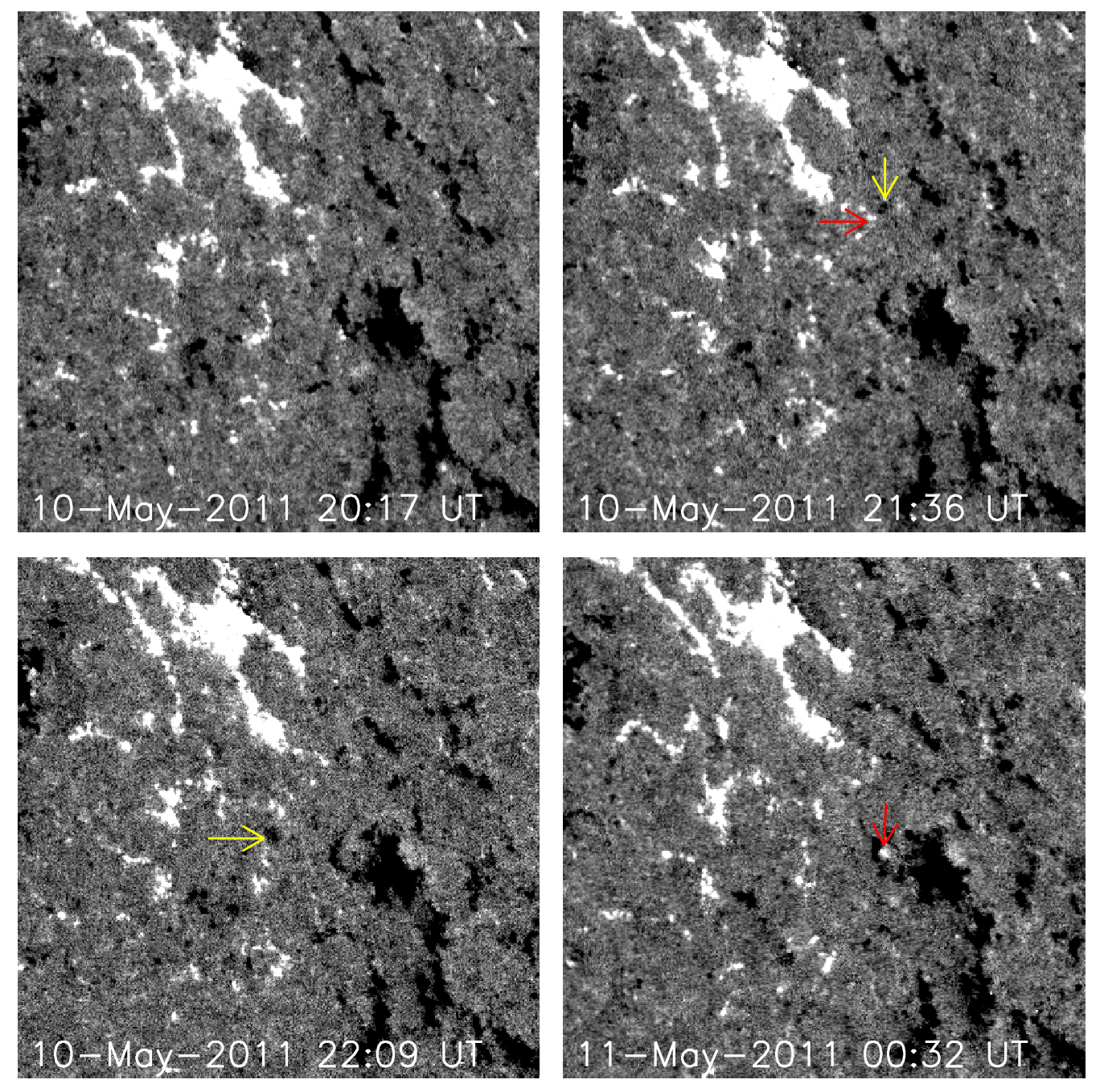}
\caption{Evolution of the photospheric magnetic field at the filament location. The red and yellow arrows indicate the sites 
of positive  and negative flux emergence.}
\label{mag}
\end{figure*}

As described above, the 2011 May 11 event was associated with the eruption of a quiescent filament. To understand the triggering of the 
filament eruption, we investigated the evolution of the photospheric magnetic field near the filament location. Some snapshots of the
magnetic field before the filament eruption are presented in Figure\ref{mag}. The filament was exactly located along a PIL. 
We observed positive and negative flux emergence at several places inside the 
filament channel before the filament eruption. The emerging positive and negative polarities are indicated by red and yellow arrows, respectively (see also the attached movie). 
Afterwards, some of them started to cancel and finally fully canceled. According to \citet{Feyn95} and \citet{Chen00}, the emergence of reconnection-favoring magnetic flux near a filament channel can trigger the eruption of the filament. 
From the above discussion, we believe that the filament eruption is due to the flux emergence and cancellation inside the filament channel. It is possible that due to this flux emergence and cancellation, the flux rope loses its equilibrium, and started to rise.

\section{Summary}
\label{discussion}

In this study, we presented the multi-wavelength observations of a GOES B-class flare occurring on 2011 May 11.
 The main results of our study are  summarized as follows:

\begin{itemize}  
\item The  observed GOES B9 flare has prolonged soft X-ray emission for more than an hour (from 02:20 UT
 to 03:20 UT) with multiple peaks. The event shows a pre-flare phase from $\sim$ 02:00 to $\sim$ 02:20 UT 
with a peak at $\sim$ 02:10 UT. 

\item The flare is also associated with almost all the other eruptive phenomena, 
such as a filament eruption, a CME, EUV waves, and type II radio bursts. Therefore this event presents an excellent 
example in which we could see all the eruptive phenomena simultaneously.

\item Before the filament eruption, we observed flux emergence and cancellation, which could be responsible for the triggering of the filament eruption.

\item The event shows that a part of the fast-mode MHD EUV wave stops at the magnetic QSL, which can be explained by mode conversion, i.e., a part of the fast-mode wave is converted to a slow-mode wave..

\item We observed the third harmonic component of the associated type II radio bursts, which is an unusual phenomenon.   

\end{itemize}

\noindent {\it Acknowledgments:}
We are grateful to the reviewers for their constructive comments and suggestions. 
The authors thank the SDO, STEREO, SOHO, HiRAS, and GONG teams for the open data policy. PFC was supported by the Chinese foundation (NSFC 11533005) and Jiangsu 333 Project.

\end{document}